\documentstyle[elsart12,osa,epsf,eqsecnum,epsfig,graphics,cite]{revtex}

\def\lsim{\raise0.3ex\hbox{$<$\kern-0.75em\raise-1.1ex\hbox{$\sim$}}}
\def\gsim{\raise0.3ex\hbox{$>$\kern-0.75em\raise-1.1ex\hbox{$\sim$}}}

\def\beq{\begin{equation}}
\def\eeq{\end{equation}}
\def\bea{\begin{eqnarray}}
\def\eea{\end{eqnarray}}

\renewcommand{\lsim}{~{\buildrel < \over {_\sim}}~}
\renewcommand{\gsim}{~{\buildrel > \over {_\sim}}~}

\newcommand{\mev}{\mathrm{MeV}}
\newcommand{\gev}{\mathrm{GeV}}
\newcommand{\tev}{\mathrm{TeV}}
\newcommand{\fm}{\mathrm{fm}}

\renewcommand{\d}{{\rm d}}

\begin{document}


\voffset1.5cm


\title{\Large\bf Testing the Color Charge and Mass Dependence of Parton Energy Loss
with Heavy-to-light Ratios at RHIC and LHC}
\author{N\'estor Armesto$^{\rm 1,2}$,
Andrea Dainese$^{\rm 3}$,
Carlos A.~Salgado$^{\rm 1}$ and\\ Urs Achim Wiedemann$^{\rm 1}$}

\address{$^{\rm 1}$ Department of Physics, CERN, Theory Division,
CH-1211 Gen\`eve 23, Switzerland\\
$^{\rm 2}$ Departamento de F\'{\i}sica de Part\'{\i}culas and Instituto Galego 
de Altas Enerx\'{\i}as, Facultade de F\'{\i}sica, Campus Sur,
Universidade de Santiago de Compostela, 15782 Santiago de Compostela,
Spain\\
$^{\rm 3}$ Universit\`a degli Studi di Padova and INFN, via Marzolo 8, 35131 Padova, Italy}
\date{\today}
\maketitle

\begin{abstract} The ratio of nuclear modification factors of high-$p_T$
heavy-flavored mesons to light-flavored hadrons (``heavy-to-light ratio'') in
nucleus-nucleus collisions tests the partonic mechanism expected to underlie
jet quenching.  Heavy-to-light ratios are mainly sensitive to the mass and
color-charge dependences of medium-induced parton energy loss.  Here, we
assess the potential for identifying these two effects in $D$ and $B$ meson
production at RHIC and at the LHC.  To this end, we supplement the
perturbative QCD factorized formalism for leading hadron production with
radiative parton energy loss. 
For $D$ meson spectra at high but experimentally accessible transverse
momentum ($10\lsim p_T \lsim 20~\gev$) in Pb--Pb collisions at the LHC, we
find that charm quarks behave essentially like light quarks. However, since
light-flavored hadron yields are dominated by gluon parents, the
heavy-to-light ratio of $D$ mesons is a sensitive probe of the color charge
dependence of parton energy loss. In contrast, due to the larger $b$ quark mass, the
medium modification of $B$ mesons in the same kinematical regime provides a
sensitive test of the mass dependence of parton energy loss.  At RHIC
energies, the strategies for identifying and disentangling the color charge
and mass dependence of parton energy loss are more involved because of the
smaller kinematical range accessible. We argue that at RHIC, the kinematical
regime best suited for such an analysis of $D$ mesons is $7 \lsim p_T \lsim
12~\gev$, whereas the study of lower transverse momenta is further complicated
due to the known dominant contribution of additional,
particle species dependent, non-perturbative effects.
\end{abstract}

\section{Introduction}
\label{sec1}

High-$p_T$ partons, produced in dense QCD matter, are expected~\cite{Gyulassy:1993hr,Baier:1996sk,Zakharov:1997uu,Wiedemann:2000za,Gyulassy:2000er,Wang:2001if} to suffer a 
significant additional medium-induced energy degradation prior to hadronization in the vacuum. 
Models based on this picture~\cite{Wang:2003aw,Dainese:2004te,Eskola:2004cr,Gyulassy:2000gk,Drees:2003zh,Vitev:2002pf,Hirano:2002sc} account for the main 
modifications of high-$p_T$ hadron production in nucleus-nucleus collisions 
at RHIC, namely the strong suppression of single inclusive hadron spectra, their centrality 
dependence~\cite{Adcox:2001jp,Adler:2003au,Adler:2002xw,Adams:2003kv,Back:2003qr,
Arsene:2003yk}, the corresponding suppression of leading back-to-back 
correlations~\cite{Adler:2002tq}, and high-$p_T$ hadron production with respect 
to the reaction plane~\cite{Adams:2004wz}. To further test the microscopic dynamics of 
medium-induced parton energy loss, two classes of measurements are now gradually coming
into experimental reach~\cite{Wiedemann:2004wp}: First, high-$p_T$ particle 
correlations~\cite{Majumder:2004wh,Majumder:2004br,Majumder:2004pt}, jet shapes and jet multiplicity
distributions~\cite{Salgado:2003rv,Pal:2003zf,Armesto:2004pt,Armesto:2004vz} will test the 
predicted relation between the energy loss of the leading 
parton, the transverse momentum broadening of the parton shower, and the softening of its 
multiplicity distribution. Second, the relative yields of identified high-$p_T$ hadrons 
will test the prediction that medium-induced parton energy loss depends on the identity
of the parent parton. Hard gluons lose more energy than hard quarks due to the stronger 
coupling to the medium~\cite{Gyulassy:1993hr,Baier:1996sk,Zakharov:1997uu,Wiedemann:2000za,Gyulassy:2000er,Wang:2001if}, and the energy loss of massive quarks is further 
reduced~\cite{Dokshitzer:2001zm,Armesto:2003jh,Zhang:2003wk,Djordjevic:2003zk} due to the 
mass-dependent restriction of the phase space into which medium-induced gluon radiation
can take place.

In the present work, we calculate the nuclear modification factor for single inclusive
high-$p_T$ spectra of charmed and beauty mesons, supplementing the perturbative 
QCD factorized formalism with radiative parton energy loss.  We also calculate the
ratio of nuclear modification factors of heavy-flavored mesons to light-flavored hadrons
(``heavy-to-light ratios''). In general, heavy-to-light ratios are sensitive to
the following medium-induced effects:
\begin{enumerate}
 \item {\it Color charge dependence of parton energy loss:} \\
 In contrast to charmed and beauty mesons, light-flavored hadron spectra receive a significant 
 $p_T$ dependent contribution from hard fragmenting gluons. Gluons are expected
 to lose more energy due to their stronger coupling to the medium. This increases
 heavy-to-light ratios at all $p_T$.
 \item {\it Mass dependence of parton energy loss:} \\
 Massive quarks are expected to lose less energy in a
 medium than light quarks.  This further enhances heavy-to-light ratios as long as
 the parton mass is not negligible compared to the partonic $p_T$. 
 \item {\it Medium-dependent trigger bias due to $p_T$ spectrum of parent parton:}\\ 
 Up to rather high transverse momentum, the 
 partonic $p_T$ spectrum of massive quarks is less steep than that of light quarks. 
 For a more steeply falling spectrum, the same parton energy loss leads to a stronger reduction 
 of the nuclear modification factor~\cite{Baier:2001yt,Eskola:2004cr}. This 
 enhances heavy-to-light ratios. 
\item {\it Medium-dependent trigger bias due to fragmentation of parent parton:}\\ 
Heavy quark fragmentation functions are
significantly harder than light quark ones. The same parton energy loss leads to a stronger
reduction of the nuclear modification factor if the fragmentation function is
harder~\cite{Dainese:2003wq}.
This reduces heavy-to-light ratios. 
\end{enumerate}
Our aim is to establish ---for the kinematical ranges accessible at RHIC and at the LHC--- the 
relative importance of these contributions to heavy-to-light ratios. In this way, we want to
assess the potential of such measurements for further clarifying the partonic mechanism
conjectured to underlie jet quenching in nucleus-nucleus collisions. The theoretical framework 
of our study is introduced in Section~\ref{sec2}, and results for the nuclear modification of heavy 
quark spectra at RHIC and at the LHC are given in Sections~\ref{sec3} and~\ref{sec4}, 
respectively. We then summarize our main conclusions.

\section{The formalism}
\label{sec2}

The nuclear modification factor $R_{AB}(p_T)$ determines the modification of the production of
a hadron $h$ in a nucleus-nucleus collisions $A$--$B$ compared to an equivalent number
of proton-proton collisions, 
\begin{equation}
R_{AB}(p_T)={\left.{\d N^{AB\to h}_{\rm medium}\over \d p_T\,\d y}\right|_{y=0} \over
\langle N^{AB}_{\rm coll}\rangle\left.{\d N^{pp\to h}_{\rm vacuum}\over \d p_T\,\d y}\right|_{y=0}}\, .
\label{2.1}
\end{equation}
Here,  $\langle N^{AB}_{\rm coll}\rangle$ is the average number of inelastic nucleon--nucleon 
collisions in a given centrality class. It is proportional to the average nuclear overlap 
function $\langle T_{AB}\rangle$, which is defined via the convolution of the nuclear thickness 
functions $T_{A,B}({\bf s})$ as an integral over the transverse plane at fixed impact
parameter ${\bf b}$, $T_{AB}(b)=\int \d{\bf s}\,T_A({\bf s})\,T_B({\bf b}-{\bf s})$.

To calculate the yield of the hadron species $h$ from a parent parton $k = q,Q,g$ (a 
massless or massive quark or a gluon) produced at rapidity $y=0$ with transverse 
momentum $p_T$, we start from a collinearly factorized expression supplemented by 
parton energy loss~\cite{Dainese:2004te,Eskola:2004cr},
\begin{eqnarray}
&& \left.{\d N^{AB\to h}_{\rm medium}\over \d p_T\,\d y}\right|_{y=0}
=  \sum_{i,j} \int \d x_i\,\d x_j\,\d (\Delta E/E)\,\d z_k\,
f_{i/A}(x_i)\,f_{j/B}(x_j) \nonumber \\
&& \qquad  \times  \left.{\d\hat N^{ij\to k}(p_{T,k}+\Delta E)\over 
\d p_{T,k}\d y}\right|_{y=0}
P(\Delta E/E,R,\omega_c,m/E)\,{D_{k\to h}(z_k)\over z_k^2}\,.
   \label{2.2}
\end{eqnarray}
Here,  $f_{i/A}(x_i)$ and $f_{j/B}(x_j)$ denote the nuclear parton distribution functions
for partons $i,j$ carrying momentum fractions $x_i$, $x_j$ in the colliding nuclei $A$, $B$, 
respectively. The total energy of the produced parton is denoted by $E$, its medium-induced 
parton energy loss by $\Delta E$. 
The produced hadron carries a fraction $z=p_{T}/p_{T,k}$ of the transverse momentum 
$p_{T,k}$ of the parent parton. The hard partonic scattering cross section 
for the production $i+j\to k +X$ reads ${\d\hat N^{ij\to k}(p_{T,k}+\Delta E)/\d p_{T,k}\d y}$. 
The fragmentation function $D_{k\to h}(z_k)$ maps the parton $k$ onto the 
hadron $h$. We work at $y=0$ where the parton energy is comparable to the parton
transverse momentum, $E\simeq p_{T,k}$. This sets 
the factorization and renormalization scales which are implicitly present in (\ref{2.2}).

The final state medium-dependence enters (\ref{2.2}) via the probability 
$P(\epsilon,R,\omega_c,m/E)$ that the parton loses an additional energy 
fraction $\epsilon = \Delta E/E$ due to medium-induced gluon radiation
prior to hadronization in the vacuum. This so-called quenching weight
depends on the in-medium path length of the hard parton and on the density
of the medium, parametrized by the variables  $R$ and $\omega_c$. 
For charm and beauty quarks, it also depends on the quark mass.
Details of the definition of the model parameters and of the calculation of $P$ are 
given in Appendix~\ref{appa} and Section~\ref{sec2b} below.

\subsection{Benchmark results without parton energy loss}
\label{sec2a}

To establish the baseline for the nuclear modification factor (\ref{2.1}),
we calculate first the high-$p_T$ particle yields (\ref{2.2}) for identified light 
and heavy-flavored hadrons in the absence of final state medium effects. 
Without medium, the quenching weight takes the form 
\begin{equation}
P(\Delta E/E,R,\omega_c,m/E)=\delta(\Delta E/E)\, , \ \ \ \ \ \ {\rm (vacuum)},
\label{2.3}
\end{equation}
and Eq.~(\ref{2.2}) reduces to the standard collinearly factorized leading order  
perturbative QCD formalism.  For this baseline, we rely on the PYTHIA event 
generator (version 6.214)~\cite{Sjostrand:2001yu}, paralleling the
approach used in Ref.~\cite{Dainese:2004te} for light-flavored hadrons.
As input, we use CTEQ~4L parton distribution functions~\cite{Lai:1999wy} with 
EKS98 nuclear corrections~\cite{Eskola:1998df}.
All partonic subprocesses $gg\to Q\overline Q$, $q\overline q\to Q\overline Q$,
$Qg \to Qg$, $\overline Q g \to \overline Q g$ and gluon
splitting $g \to Q \overline Q$ are included in PYTHIA (option $\rm MSEL=1$). 
PYTHIA also accounts for the possibility that a $Q\overline Q$-pair
is created by (vacuum) gluon radiation from the primary partons created in 
the hard collision. This effect becomes significant at the LHC but is negligible
at RHIC. The splitting $g \to Q\overline Q$ requires a high virtuality gluon 
($\gsim 2\,m_Q$) which corresponds to a short formation time. Hence, we 
shall assume later that heavy quark pairs from
such a secondary $g \to Q\overline Q$ production process have the 
same in-medium path length as those produced directly in the hard scattering process.
\vspace{-0.5cm}
\begin{figure}[t]\epsfxsize=8.7cm \epsfysize=15.7cm
\centerline{\epsfbox{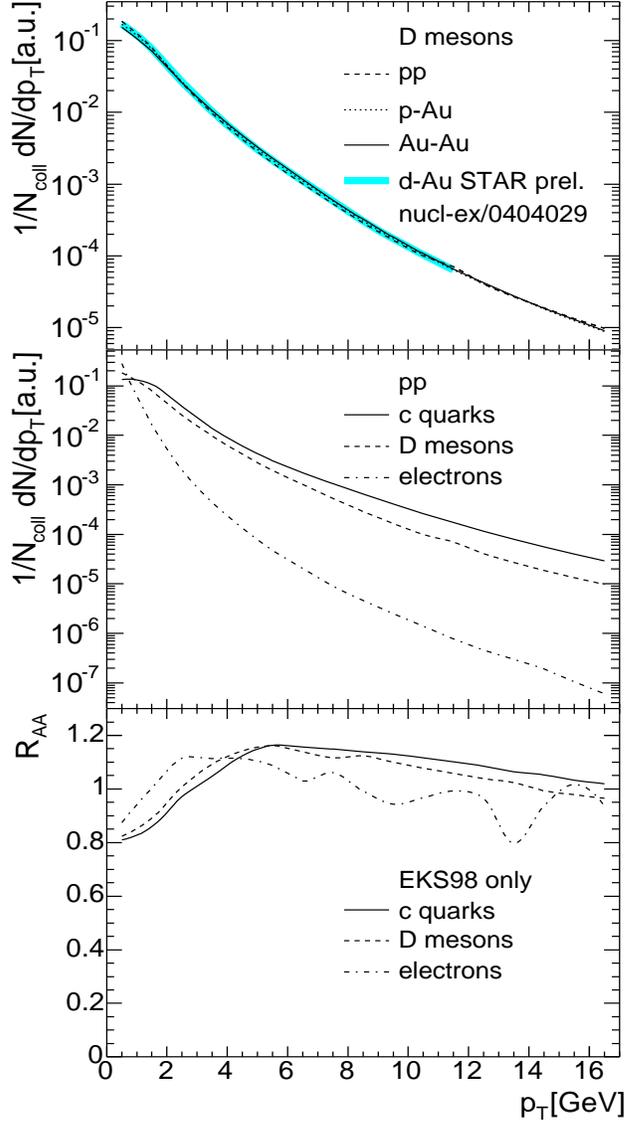}}
\vspace{0.5cm}
\caption{Upper and central panels: transverse momentum spectra
for $D$ mesons, $c$ quarks and electrons from charm decays
in pp, p--Au and Au--Au collisions at $\sqrt{s_{NN}}=200~\gev$. 
The shaded band in the
upper panel corresponds to a parameterization of data from the STAR
Collaboration~\cite{Tai:2004bf,Adams:2004fc}.
Lower panel: nuclear modification factor for $c$ quarks, $D$ mesons
and electrons from charm decay in Au--Au collisions at 
$\sqrt{s_{NN}}=200~\gev$. Except for the nuclear modification of parton 
distribution functions according to EKS98~\cite{Eskola:1998df}, 
no medium effects have been included.
}\label{fig1}
\end{figure}

Our parameterization of the fragmentation functions in (\ref{2.2}) is based on
the string model implemented in PYTHIA. In particular, we generate with 
PYTHIA pp events which contain charm (or beauty) quarks and we force the 
semi-electronic decay of corresponding heavy-flavored mesons. From these 
events we extract the heavy quark yields and the probabilities for a
quark $k$ of transverse momentum $p_{T,k}$ to fragment into a hadron 
$h$ with $p_{T}$ [and further into an electron with $p_{T,e}$]. These probabilities will 
be used as fragmentation functions. 

For RHIC energy, $\sqrt{s_{NN}}=200~\gev$, charm production in PYTHIA was 
tuned to reproduce, in shape, the experimental data on the $D$ meson $p_T$ 
distribution in d--Au collisions measured
by the STAR Collaboration~\cite{Tai:2004bf,Adams:2004fc}.
This baseline for charm production at RHIC is presented in Fig.~\ref{fig1}.
The $p_T$ distribution of $D$ mesons traces closely that of their parent
$c$ quarks, but the distribution of electrons is considerably softer. This
complicates attempts to study heavy quark parton energy loss 
on the basis of single inclusive electron
spectra~\cite{Batsouli:2002qf,Lin:2004dk}. Nuclear modifications of the
parton distribution functions are seen to affect the $p_T$ spectra at most by  
$\pm 20\%$ at low $p_T$.

For LHC energy, $\sqrt{s_{NN}}=5.5~\tev$, 
we use a tuning of PYTHIA~\cite{Carrer:2003ac} that reproduces the shape of 
the $p_T$ distributions for charm and beauty quarks given by pQCD 
calculations at next-to-leading order~\cite{Mangano:1991jk} with CTEQ~4M 
parton distribution functions, $m_c=1.2~\gev$ and factorization and 
renormalization scales $\mu_F=\mu_R=2\,m_T$ for charm, and 
$m_b=4.75~\gev$ and $\mu_F=\mu_R=m_T$ for beauty, where 
$m_T^2=m_Q^2+p_T^2$. 

\subsection{Modeling the medium dependence of parton energy loss}
\label{sec2b}

Medium-induced parton energy loss depends on the in-medium
path length and the density of the medium. It is characterized by the medium-induced
gluon energy distribution $\omega {\it d}I^{\rm med}/{\it d}\omega$ radiated off the hard 
parton. This defines the probability distribution for medium-induced
energy loss (``quenching weight $P$''), entering the cross section (\ref{2.2}) 
for medium-modified high-$p_T$ hadron production.  Further details of this
procedure are given in Appendix~\ref{appa}. 

The dependence of parton energy loss
on density and in-medium path length can be characterized
in terms of the time-dependent BDMPS transport coefficient $\hat{q}(\xi)$ which
denotes the average squared transverse momentum transferred from the medium to
the hard parton per unit path length. Numerically, one finds that the effects of a 
time-dependent density of the medium on parton energy loss can be accounted for by an
equivalent static medium, specified in terms of the time-averaged model parameters
$\omega_c$ and $R$,
\begin{eqnarray}
        \omega_c &\equiv&  \int_0^\infty \d\xi\,\xi\,\hat{q}(\xi)\, ,
        \label{2.4}\\
        R&=&{2 \omega_c^2\over \int_0^\infty \d\xi\,\hat{q}(\xi)}\, .
         \label{2.5}
\end{eqnarray}
For light quarks and gluons, the quenching weights $P(\Delta E /E, R, \omega_c)$
have been calculated in Ref.~\cite{Salgado:2003gb} and they are available as a 
CPU-inexpensive FORTRAN routine. For the purpose of this work, we have 
extended this calculation to the case of quenching weights 
$P(\Delta E /E, R, \omega_c,m/E)$ for massive quarks, starting from the 
medium-induced gluon energy distribution determined in Ref.~\cite{Armesto:2003jh}. 
Results for these quenching weights are given in Appendix~\ref{appa} and they are
publicly available as a CPU-inexpensive FORTRAN routine
accompanying this paper~\cite{ourref}. 

We sample the positions ${\bf s}=(x_0,y_0)$ of the parton production 
points in the transverse plane of a nucleus-nucleus collision $A$--$B$ with the probability
given by  the product of the nuclear thickness functions $T_A({\bf s})\, T_B({\bf b-s})$. 
For these thickness functions, we choose Woods-Saxon parameterizations of nuclear density profiles~\cite{DeJager:1974dg}. For a hard parton with production point ${\bf s}=(x_0,y_0)$ and 
azimuthal propagation direction ${\bf n}=(n_x,n_y)$, the local transport coefficient along the path 
of the parton is defined as~\cite{Dainese:2004te}:
\begin{equation}
\hat{q}(\xi)=k\,
T_A( {\bf s}+  \xi {\bf n})\,T_B({\bf b}-[{\bf s}+  \xi {\bf n}])\, .
\label{2.6}
\end{equation}
Here, the constant $k$ sets the scale of the transport coefficient. 
This defines $\omega_c$ and $R$ in (\ref{2.4}) and (\ref{2.5}).
All values of transport coefficients used in this work characterize time-averaged 
properties of the medium ---their numerical value is determined by the
established relation~\cite{Salgado:2003gb} between parton energy 
loss in a dynamically expanding and a static medium.

Partons that 
lose their entire energy due to medium effects are redistributed in our formalism
according to a thermal distribution~\cite{Lin:1997cn},
\begin{equation}
\left.{\d N_{\rm thermal}\over \d m_T\,\d y}\right|_{y=0}\propto m_T\,
\exp{\left(-{m_T\over T}\right)}\,.
\label{2.7}
\end{equation}
For the following results, we use $T=300~\mev$. By varying $T$ between 5 and 
$500~\mev$, we have checked that the choice of $T$ affects the nuclear modification 
factor (\ref{2.1}) only for  $p_T \lsim 3~\gev$. The reasons why the present parton
energy loss formalism is not reliable at such low transverse momenta have been 
mentioned repeatedly~\cite{Salgado:2003gb,Wiedemann:2004wp}. Accordingly, 
the main conclusions drawn from our study will be for significantly higher transverse 
momentum.

\begin{figure}[t]\epsfxsize=12.0cm
\centerline{\epsfbox{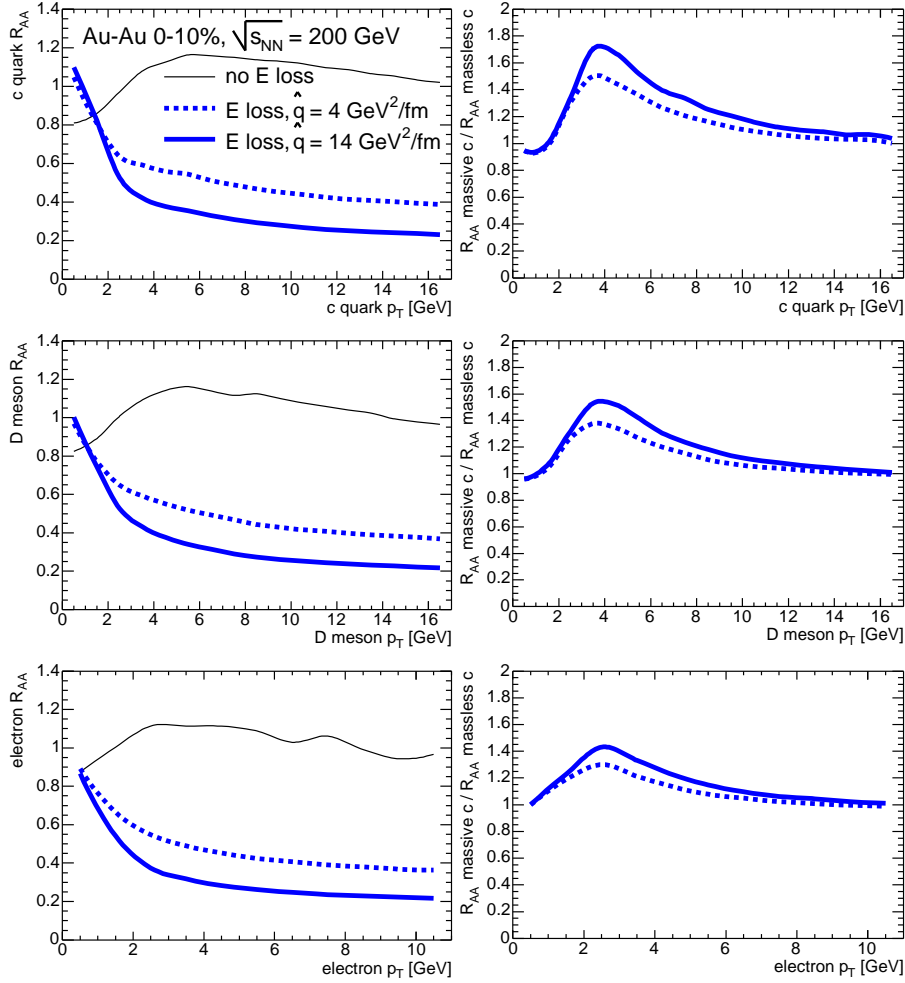}}
\vspace{0.5cm}
\caption{LHS: Nuclear modification factors for $c$ quarks, $D$ mesons and 
their decay electrons in central (0--10\%) Au--Au collisions at $\sqrt{s_{NN}}=200~\gev$
for different time-averaged strengths of the parton energy loss. RHS: The ratio of the
nuclear modification factor plotted on the LHS, divided by the same factor calculated
for a massless charm quark.
}\label{fig2}
\end{figure}

\section{Results for RHIC}
\label{sec3}

Recent model studies of the nuclear modification of light-flavored hadrons in Au--Au 
collisions at RHIC favor a time-averaged transport coefficient  
$\hat{q} \approx 4\div 14~{\rm GeV}^2/{\rm fm}$~\cite{Dainese:2004te,Eskola:2004cr}. 
To account for the significant 
systematic uncertainties of this favored value, we calculate here the particle yield (\ref{2.2}) 
for $\hat{q} = 0$, $4$ and
$14\, {\rm GeV}^2/{\rm fm}$. The corresponding nuclear modification 
factors are shown in Fig.~\ref{fig2} for $D$ mesons,  $c$ quarks and for electrons from 
charm decay. To illustrate the mass dependence of parton energy loss, we compare the
calculation for a realistic charm quark mass $m_c=1.2~\gev$ to 
the hypothetical case that the charm quark loses as much energy
as a light quark. The ratio of the two nuclear modification factors thus obtained is
shown in Fig.~\ref{fig2}. One sees that the mass dependence of final state parton
energy loss leads to a significant change of the nuclear modification factor of $D$ mesons 
up to transverse momenta of $p_T \lsim 12~\gev$. This is consistent with the general
observation that the mass dependence of parton energy loss is a function of
$m/E$ and becomes negligible for $m/E < 0.1$~\cite{Armesto:2003jh}. 

In Fig.~\ref{fig2}, one also sees that the ratio of the realistic nuclear modification factor 
to the one for $m_c=0~\gev$ has a maximum around $p_T \sim 2$--$4~\gev$, and drops 
slightly below unity for very small $p_T$. This is a generic consequence of the fact that parton 
energy loss redistributes charm quarks towards the softer region in transverse phase 
space. If the depletion at high $p_T$ is less significant due to a finite parton mass, 
then the resulting enhancement at small $p_T$ is also less significant and this 
depletes the ratio of nuclear modification factors. However, at smaller transverse 
momentum, $p_T < 7~\gev$,
soft hadron production or non-perturbative hadronization mechanisms in the medium like
recombination or
coalescence~\cite{Gorenstein:2000ck,Andronic:2003zv,Lin:2003jy}
(and the possibility of
thermalization~\cite{vanHees:2004gq,Moore:2004tg} and collisional
energy loss~\cite{Romatschke:2004au,Mustafa:2004dr})
have to be considered~\cite{Wiedemann:2004wp} to account for 
the sizable particle species dependence of the nuclear modification 
factors~\cite{Adams:2003am,Adler:2003kg}. 
Here, parton energy loss alone cannot be expected to provide a reliable description.
This complicates the analysis of heavy meson spectra and their decay products
at low $p_T$~\cite{Adler:2004ta}.
As a consequence, the following discussion will mainly focus on the region $p_T > 7~\gev$.

\begin{figure}[t]\epsfxsize=11.7cm
\centerline{\epsfbox{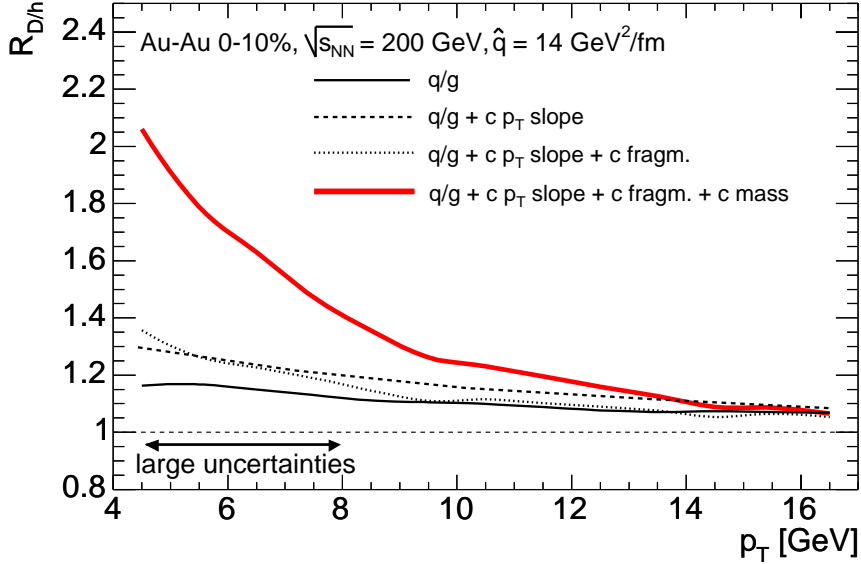}}
\vspace{0.5cm}
\caption{Different contributions to the heavy-to-light ratio of $D$ mesons 
in central (0--10\%) Au--Au collisions at $\sqrt{s_{NN}}=200~\gev$. Different
curves correspond to the case in which the charm distribution is described
i) with the same $p_T$ spectrum, fragmentation function and parton energy loss
as a light quark, ii) with a realistic charm $p_T$ spectrum only, iii) with the
charm $p_T$ spectrum and fragmentation function of a realistic charm quark
and iv) for the realistic case which includes the mass dependence of parton
energy loss.
}\label{fig3}
\end{figure}

As explained in the Introduction, the mass dependence of parton energy loss
displayed in Fig.~\ref{fig2} is only one of several parton species dependent
modifications of parton energy loss. Other modifications result from the dependence
of parton energy loss on the color charge of  the parent parton, and from trigger bias
effects related to the partonic $p_T$ spectrum and the parton fragmentation function.
To disentangle the relative strength of these effects, we plot  in Fig.~\ref{fig3}  the 
heavy-to-light ratio $R_{D/h}=R^D_{AA}/R^h_{AA}$ ($h$ referring to light-flavored hadrons)
for model calculations in which the above mentioned mass-sensitive medium dependences
have been switched off selectively. 

Parametrically, the mass dependence of the medium-induced parton energy 
loss and of the trigger biases becomes negligible at high transverse momentum 
where $m_c/p_T \to 0$. In contrast, the difference between quark and gluon energy 
loss stays at all $p_T$ due to the ratio of the Casimir factors $C_A/C_F=9/4$. Hence, 
at the highest $p_T$, the color charge dependence of parton energy loss dominates 
the difference between the nuclear modification factor for light-flavored and 
heavy-flavored hadrons. In agreement
with this argument, Fig.~\ref{fig3} shows that the color charge dependence accounts for
the dominant deviation of $R_{D/h}$ from unity for $p_T \gsim 12~\gev$. At
such high transverse momentum, charm quarks start to behave 
like light quarks and the heavy-to-light ratio of $D$ mesons becomes mainly sensitive
to the color charge dependence of parton energy loss. However, at RHIC energies,
the corresponding signal is rather small, see Fig.~\ref{fig3}.
In combination with the small high-$p_T$ cross sections, this may limit a quantitative
study of this interesting kinematical regime.

Medium-dependent trigger bias effects due to the mass dependence of the partonic 
$p_T$ spectrum and due to the fragmentation function largely compensate each other. 
They go in the directions indicated by the general arguments in the Introduction,
but they are negligible over the entire $p_T$ range displayed in Fig.~\ref{fig3}.

The mass dependence of parton energy loss dominates the deviation of
$R_{D/h}$ from unity for $p_T < 12~\gev$. However, as discussed above, 
the particle species dependence of $R_{AA}$ 
for $p_T \lsim 7~\gev$ ~\cite{Adams:2003am,Adler:2003kg} 
makes the application of parton energy loss questionable.
Thus, to assess the mass dependence of parton energy loss with the
heavy-to-light ratio $R_{D/h}$ at RHIC, one should focus on the kinematical 
region, $7 \lsim p_T \lsim 12~\gev$. Even in 
that region, the color charge effect is sizable and has to be taken into account
in a quantitative analysis, see Fig.~\ref{fig3}. 

\section{Results for the LHC}
\label{sec4}

To calculate the nuclear modification of high-$p_T$ particle yields in Pb--Pb collisions 
at LHC energy $\sqrt{s_{NN}}=5500~\gev$, the density of the produced matter has
to be characterized, e.g. by  the BDMPS transport coefficient $\hat q$,
see Section~\ref{sec2}.  This transport coefficient is proportional to the particle multiplicity 
in the collision. In Refs.~\cite{Dainese:2004te,Eskola:2004cr} the relative increase of the event
multiplicity from RHIC to LHC has been taken to be $\sim 7$~\cite{Eskola:1999fc}, 
but other more recent estimates give a significantly smaller increase $\sim 2.6$~\cite{Armesto:2004ud}. 
Here, we scan a very wide range of the model parameter space by varying $\hat q$
between a low estimate at RHIC energies and the highest estimates for LHC energies,
$\hat q=4$,  $25$ and $100~\gev^2/\fm$.

\begin{figure}[t]\epsfxsize=11.7cm
\centerline{\epsfbox{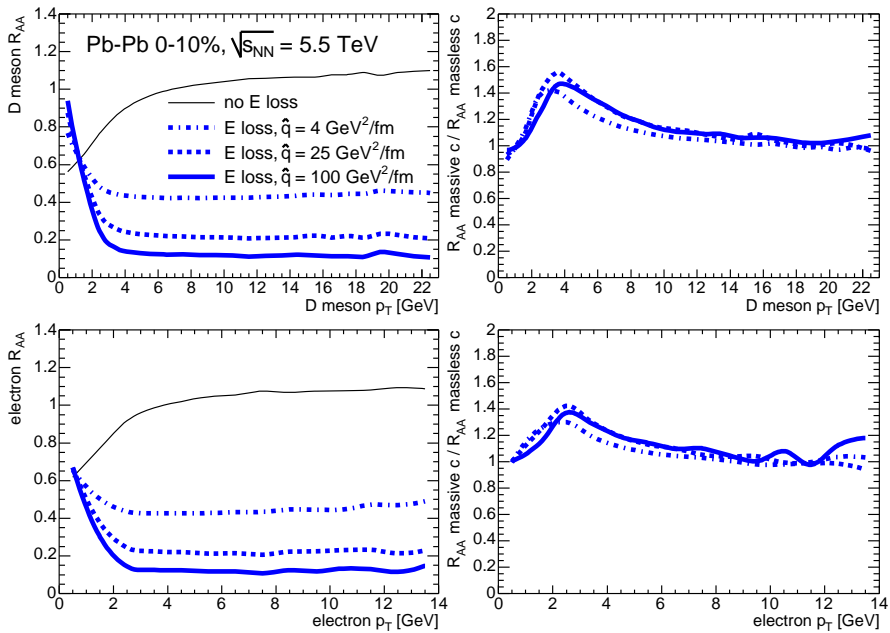}}
\vspace{0.5cm}
\caption{LHS: Nuclear modification factors for $D$ mesons (upper plot) and electrons 
from charm decays (lower plot) in central (0--10\%) Pb--Pb collisions at
$\sqrt{s_{NN}}=5.5~\tev$. RHS: The ratio of the realistic nuclear modification factors shown
on the left hand side and the same factors calculated by solely neglecting the mass
dependence of parton energy loss. 
}\label{fig6}
\end{figure}

At the LHC as at RHIC, a significant mass dependence of the nuclear modification factor 
of $D$ mesons and of their decay electrons is limited to transverse momenta below 
$p_T \lsim 10~\gev$, see Fig.~\ref{fig6}. However, at the LHC, there are arguments
that non-perturbative hadronization mechanisms such as recombination may dominate 
the medium modification of identified particle yields up to even higher transverse momenta 
than at RHIC~\cite{Accardi:2003gp}. Thus, the mass dependence of parton energy loss will dominate
the deviation of the heavy-to-light ratio of $D$ mesons from unity only in a rather small
kinematical window, if at all.

\begin{figure}[t]\epsfxsize=11.7cm
\centerline{\epsfbox{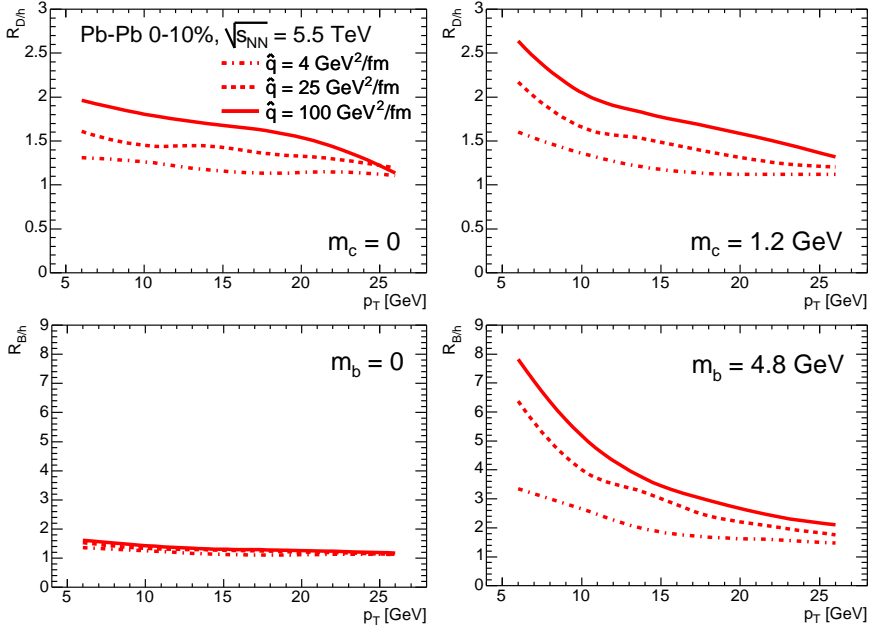}}
\vspace{0.5cm}
\caption{Heavy-to-light ratios for $D$ mesons (upper plots) and $B$ mesons (lower
plots) for the case of a realistic heavy quark mass (plots on the right) and for a case
study in which the quark mass dependence of parton energy loss is neglected
(plots on the left).
}\label{fig8}
\end{figure}

At higher transverse momenta, $p_T \gsim 10~\gev$, the charm mass dependence of 
parton energy loss becomes negligible since $m_c/p_T \to 0$, see Fig.~\ref{fig6}. 
Remarkably, however, even if charm mass effects are neglected, the
heavy-to-light ratio $R_{D/h}$ shows for realistic  model parameters a
significant enhancement $R_{D/h} \sim 1.5$ in a theoretically rather clean 
and experimentally accessible kinematical regime of high transverse momenta 
$10 \lsim p_T \lsim 20~\gev$, see Fig.~\ref{fig8} (upper panels). The
reason is that parton production at mid-rapidity tests values of Bjorken $x$ which are a 
factor $\sim 30$ smaller at LHC than at RHIC. At smaller Bjorken $x$, a larger fraction 
of the produced light-flavored hadrons have gluon parents and thus the color charge dependence
of parton energy loss can leave a much more sizable effect in the
heavy-to-light ratio $R_{D/h}$ at LHC. In summary, charm quarks 
giving rise to $D$ mesons in the kinematical range $10 \lsim p_T \lsim 20~\gev$
behave essentially like massless quarks in Pb--Pb collisions at the LHC. But the significant
gluonic contribution to light-flavored hadron spectra in this kinematical range makes the
heavy-to-light ratio $R_{D/h}$ a very sensitive hard probe for testing the color charge 
dependence of parton energy loss. 

\begin{figure}[t]\epsfxsize=11.7cm
\centerline{\epsfbox{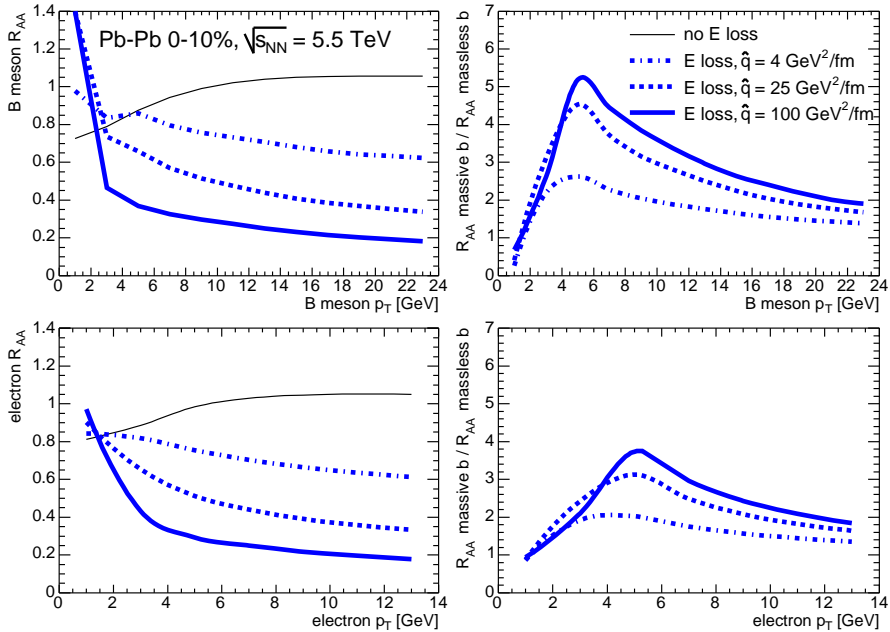}}
\vspace{0.5cm}
\caption{The same as Fig.~\protect{\ref{fig6}} but for $B$ mesons and electrons
from beauty decays.
}\label{fig7}
\end{figure}

At the higher LHC energies, the higher mass scale of $b$ quarks can be 
tested in the corresponding nuclear modification factors and heavy-to-light ratios 
for $B$ mesons and for electrons from $b$ decays. As seen in Figs.~\ref{fig8} and~\ref{fig7}, 
for transverse momenta $10 \lsim p_T \lsim 20~\gev$, the mass 
dependence of parton energy loss modifies the nuclear modification factor by a 
factor 2 or more. It dominates over the color charge dependence. As for all spectra 
discussed above, the medium-dependence of trigger
bias effects is rather small for beauty production at the LHC [In Fig.~\ref{fig8}, these
trigger bias effects account for the small but visible differences between
$R_{D/h}$ and $R_{B/h}$ in the model calculation in which the mass dependence of
parton energy loss has been neglected.]. We conclude that the heavy-to-light ratio 
$R_{B/h}$ in Pb--Pb collisions at the LHC provides a very sensitive hard probe for 
testing the parton mass dependence of parton energy loss in the theoretically rather 
clean and experimentally accessible kinematical regime of 
$10 \lsim p_T \lsim 20~\gev$.

\section{Conclusions}
\label{sec5}

The mass dependence of parton energy loss and its phenomenological consequences 
for heavy-to-light ratios have received significant interest recently
\cite{Dokshitzer:2001zm,Armesto:2003jh,Zhang:2003wk,Djordjevic:2003zk,Lokhtin:2004rg,Zhang:2004qm,Djordjevic:2004nq,Xiang:2005et,Thomas:2004ie}, since they provide a tool 
to test the conjectured microscopic dynamics underlying the phenomenon of high-$p_T$ 
hadron suppression at RHIC. Here, we have shown that heavy-to-light ratios are not
solely sensitive to the mass dependence of parton energy loss, but also to its color 
charge dependence. The task for both experiment and theory is to identify and to 
disentangle both effects. This issue is of fundamental importance since the more precise 
understanding of the dynamics modifying a hard probe is a prerequisite for the more 
precise understanding of the properties of what is probed, namely dense QCD matter.

Previous proposals to search for the color charge dependence of parton energy loss 
either tried to exploit the weak $p_T$ dependence of the quark-to-gluon ratio of parent
partons in light-flavored hadron spectra~\cite{Wang:2004tt}. Alternatively, they focused on 
proton-to-antiproton ratios at high $p_T$~\cite{Wang:1998bh}. Since antibaryons receive 
a larger fragmentation contribution from gluon parents, parton energy loss predicts an 
increase of the $p/\overline{p}$ ratio with increasing $p_T$. However, the kinematical
range for studying proton-to-antiproton ratios is rather restricted due to experimental 
limitations on measuring identified baryons at high-$p_T$, and due to the anomalous 
baryon-to-meson ratio which complicates the physics interpretation at intermediate 
$p_T$ ($p_T \lsim 7~\gev$).  Similar caveats apply to  the ratio $\Lambda/\overline{\Lambda}$. 
This baryon-to-antibaryon ratio may have the experimental advantage that it can be 
measured via topological reconstruction to higher $p_T$ than the $p/\overline{p}$ ratio.
However, its sensitivity to the medium-modification of parton dynamics remains to be 
assessed, and may be complicated due to our limited knowledge about $\Lambda$ 
fragmentation functions. Compared to these measurements, the heavy-to-light ratios
of $D$ mesons have significant experimental and theoretical advantages. In particular,
in Pb--Pb collisions at the LHC, they show a high sensitivity to the color charge 
dependence of parton energy loss in a kinematical range $10 \lsim p_T \lsim 20~\gev$ 
in which charm quarks behave like light quarks and other medium modifications are 
expected to be negligible, see Section~\ref{sec4}. Thus, they provide 
a unique way to ``tag'' a light quark (namely the charm quark) since 
they can be characterized experimentally by cross-checking several 
relatively clean decay channels. 

On the other hand, the heavy-to-light ratio of $B$ mesons in the same kinematical regime 
$10 \lsim p_T \lsim 20~\gev$ at the LHC is expected to be predominantly sensitive
to the mass dependence of parton energy loss, showing significant enhancement
factors of the order $2\div 5$, see Section~\ref{sec4}. Thus, a combined analysis of open
charm and beauty mesons at the LHC provides the means to quantify and to disentangle 
the characteristic differences of the strength of the parton energy loss for the different 
identities of the parent parton.

In nucleus-nucleus collisions at RHIC, the task of identifying and disentangling
the color charge and mass dependence of parton energy loss is more challenging, 
since a significantly smaller kinematical range of high transverse momentum is experimentally
accessible. In particular, the region up to at least $p_T \lsim 7~\gev$ can only
provide circumstantial evidence to this end, since it is known to receive significant 
additional particle species dependent non-perturbative contributions. In view of
the present study, the first use of RHIC heavy meson spectra in the context of testing
parton energy loss is to test the {\it combined} effect of the mass and color charge 
dependence of parton energy loss in the kinematical
range $7 \lsim p_T \lsim 12~\gev$ where $R_{D/h}$ is expected to be gently but
visibly enhanced above unity, see Section~\ref{sec3}. 

We finally note that correlation measurements may provide important complementary 
information for elucidating the influence of parton identity on final state parton energy loss.
For example, requiring that a high-$p_T$ trigger hadron at forward rapidity is balanced
by a recoil at mid-rapidity, one may be able to study medium-modified hadron production
in a configuration which enriches the contribution of gluon parents. Both at RHIC and at
the LHC, the class of 
correlation measurements with this potential is large. It includes many as yet unexplored 
observables such as three jet events which in large acceptance experiments at the LHC 
may give access to the study of well-separated samples of quark and gluon jets.
These correlations lie outside the scope of the present work. To the best of our
knowledge, it is an open question whether some of them have a similar 
or even higher sensitivity to the mass and color charge dependence of parton energy 
loss than the ratios of particle identified single inclusive hadron spectra studied here.

{\bf Acknowledgment:} We thank Peter Jacobs for helpful discussions.  

\appendix

\section{Quenching weights for massive partons} 
\label{appa}

In this appendix, we give details of how to calculate the probability 
$P(\Delta E/\omega_c,R,m/E)$ that a quark of mass $m$ and initial energy $E$ 
loses an energy fraction $\Delta E / E$ due to medium-induced gluon radiation. 

We start from the medium-induced distribution of gluons
of energy $\omega$ and transverse momentum ${\bf k}_\perp$, radiated off the 
hard massive quarks due to multiple scattering in the spatially extended 
medium~\cite{Armesto:2003jh}
\begin{eqnarray}
  \omega\frac{dI^{\rm med}}{d\omega\, d{\bf k}_\perp}
  &=& {\alpha_s\,  C_F\over (2\pi)^2\, \omega^2}\,
    2{\rm Re} \int_{0}^{\infty}\hspace{-0.3cm} \d y_l
  \int_{y_l}^{\infty} \hspace{-0.3cm} \d\bar{y}_l\,
  e^{i \bar{q} (y_l - \bar{y}_l)}
   \int \d{\bf u}\, e^{-i{\bf k}_\perp \cdot {\bf u}}   \,
  e^{ -\frac{1}{2} \int_{\bar{y}_l}^{\infty} \d\xi\, n(\xi)\,
    \sigma({\bf u}) }\,
  \nonumber \\
  && \times {\partial \over \partial {\bf y}}\cdot
  {\partial \over \partial {\bf u}}\,
  \int_{{\bf y}=0={\bf r}(y_l)}^{{\bf u}={\bf r}(\bar{y}_l)}
  \hspace{-0.5cm} {\cal D}{\bf r}
   \exp\left[ i \int_{y_l}^{\bar{y}_l} \hspace{-0.2cm} \d\xi
        \frac{\omega}{2} \left(\dot{\bf r}^2
          - \frac{n(\xi) \sigma\left({\bf r}\right)}{i\, \omega} \right)
                      \right]\, .
    \label{a.1}
\end{eqnarray}
The strong coupling constant $\alpha_s$ and Casimir operator $C_F = \frac{4}{3}$ 
determine the coupling strength of gluons to the massive quark. The physical interpretation 
of the internal integration variables in (\ref{a.1}) has been explained 
elsewhere~\cite{Salgado:2003gb,Armesto:2003jh} and plays no role in what follows. 
For numerical calculations, we use $\alpha_s = 1/3$.
Eq.~(\ref{a.1}) resums the effects of arbitrary many medium-induced scatterings 
to leading order in $1/E$. 

The parton mass dependence enters the gluon energy distribution (\ref{a.1}) via the 
phase factor $\exp \left[i \bar{q} (y_l - \bar{y}_l)\right]$~\cite{Armesto:2003jh},
where $\bar{q}$ is defined as the difference between the total
three momentum of the initial quark ($p_1$), and the final
quark ($p_2$) and gluon ($k$),
\begin{equation}
  \bar{q} = p_1 - p_2 - k \simeq \frac{x^2\, m^2}{2 \omega}\, ,
  \qquad x = \frac{\omega}{E}\, .
  \label{a.2}
\end{equation}
%

\begin{figure}[htb]
\begin{center}
\epsfig{file=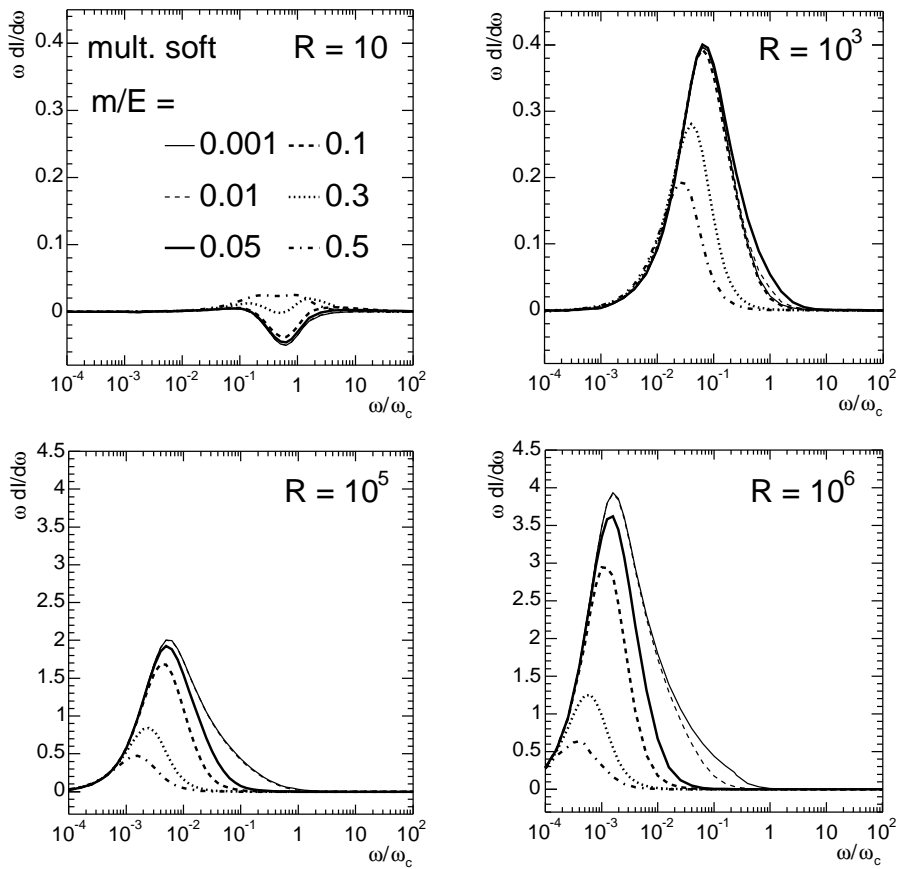,width=12.5cm}
\caption{The $k_T$ integrated medium-induced gluon energy distribution (\ref{a.1})
radiated off a massive quark for different values of $R$ (panels) 
and of $m/E$ (line styles).}
\label{figa1}
\end{center}
\end{figure}

The medium-dependence enters (\ref{a.1}) via the product of the 
time-dependent density $n(\xi)$ of scattering centers times the strength of a single 
elastic scattering $\sigma({\bf r})$. In what follows, we work in the multiple soft scattering
approximation 
\begin{equation}
        n(\xi)\sigma({\bf r}) \simeq \frac{1}{2} \hat q(\xi)r^2\, ,
        \label{a.3}
\end{equation}
where the path integral in (\ref{a.1}) can be evaluated in a saddle point approximation. 
This approximation is known to lead to results which are physically 
equivalent~\cite{Salgado:2003gb} to other approaches. For a hard parton which
transverses a time-independent medium of length $L$, we have
$\hat{q}(\xi) = \hat{q}_0\, \Theta(L-\xi)$. For the realistic case of an expanding 
medium~\cite{Gyulassy:2000gk,Salgado:2002cd,Baier:1998yf},  the radiation spectrum is the same as that 
for an equivalent static medium of appropriately
rescaled transport coefficient. This dynamical scaling law is used to define in
(\ref{2.4}) and (\ref{2.5}) the only medium-dependent parameters $\omega_c$ and  
$R$.  In Fig.~\ref{figa1} the medium-modified part of the ${\bf k}_\perp$ integrated 
gluon energy distribution (\ref{a.1}) is plotted for different
values of $m/E$ and $R=\omega_c L$ where $\omega_c=\frac{1}{2}{\hat q}\,L^2$.

\begin{figure}[h]\epsfxsize=10.0cm
\centerline{\epsfbox{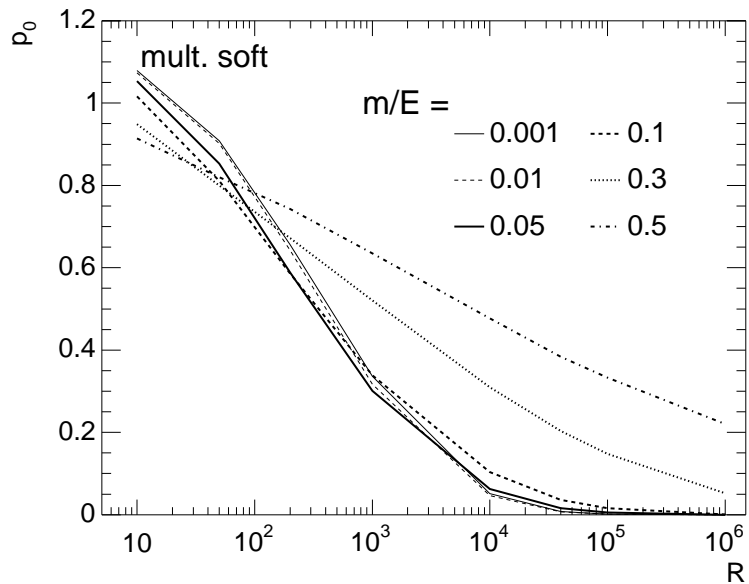}}
\vspace{0.5cm}
\caption{Discrete part $p_0$ of the quenching weight (\protect\ref{a.4}) as a function of $R$ 
for different  values of $m/E$.
}\label{figa2}
\end{figure}

The probability that in the parton fragmentation process of the hard massive quark,
an additional amount $\Delta E$ of the initial quark energy is lost due to
multiple medium-induced gluon radiation, can be modeled by a Poissonian 
process~\cite{Baier:2001yt},
\begin{eqnarray}
  P(\Delta E/\omega_c,R,m/E) &=& \sum_{n=0}^\infty \frac{1}{n!}
  \left[ \prod_{i=1}^n \int \d\omega_i \frac{dI^{\rm med}(\omega_i)}{d\omega}
    \right]
\label{a.4}\\
&& \quad \times 
    \delta\left(\Delta E -\sum_{i=1}^n {\omega_i} \right)
    \exp\left[ - \int \d\omega \frac{dI^{\rm med}}{d\omega}\right]\ . \nonumber
\end{eqnarray}
We calculate this probability distribution via its Mellin transform as described
in Ref.~\cite{Baier:2001yt,Salgado:2003gb}. It has a discrete and a continuous part 
\begin{equation}
  P(\Delta E/E,R,\omega_c)
  =p_0(R,m/E)\delta(\Delta E/\omega_c)+p(\Delta E/\omega_c,R,m/E)\, .
  \label{a.5}
\end{equation}
The discrete contribution $p_0$ is plotted in Fig.~\ref{figa2}. It denotes the probability 
that the hard parton escapes the collision region without further interaction~\cite{Salgado:2002cd}. Accordingly, this probability decreases with increasing density or increasing in-medium 
path length. For sufficiently large in-medium path length or density, $R = \omega_c\, L > 100$, 
the discrete part $p_0$ also increases with $m/E$ ---this is consistent with a mass-dependent
reduction of parton energy loss. 

\begin{figure}[h]\epsfxsize=10.0cm
\centerline{\epsfbox{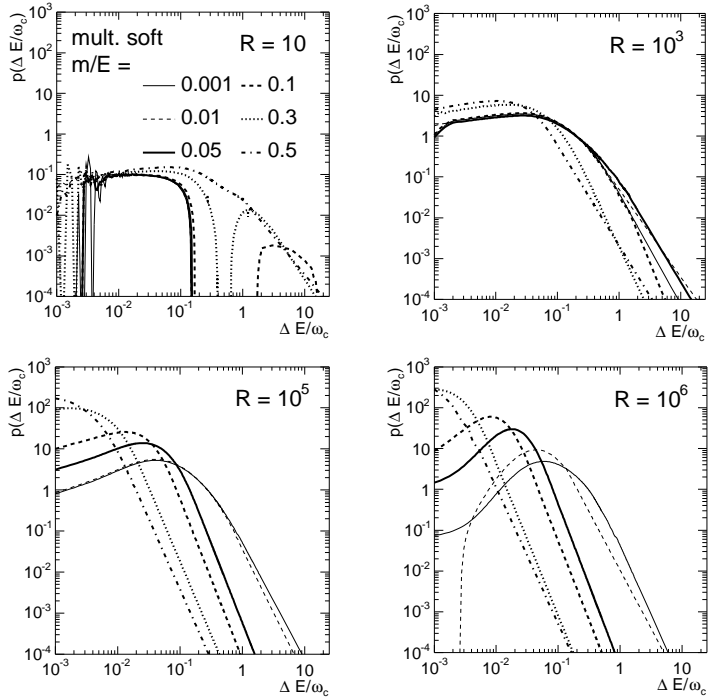}}
\vspace{0.5cm}
\caption{Continuous part of the quenching weight (\protect\ref{a.4})
for different values of $R$ and of $m/E$.
}\label{figa3}
\end{figure}

The continuous part of (\ref{a.5}) denotes the probability that the hard parton loses an 
additional energy $\Delta E$ due to medium-induced gluon radiation. As seen in Fig.~\ref{figa3},
this part increases with increasing in-medium path length or increasing density, since it
depends on $\Delta E/(\hat{q}L^2/2)=\Delta E/\omega_c$. Also, for sufficiently large 
$R \gsim 1000$, the probability of losing a large energy $\Delta E$ is seen to decrease 
with increasing mass, as expected for a mass-dependent reduction of parton energy loss. 
In contrast, for the parameter range $R < 1000$, the relation between parton energy loss 
and mass dependence is more complicated and not monotonic, see Fig.~\ref{figa3} for details. 
The latter case is of little practical relevance since it corresponds to very small medium 
effects, see also~\cite{Salgado:2003gb}.

The quenching weight $P(\Delta E/\omega_c,R,m/E)$ is a generalized probability
which is normalized to unity. We have checked that the numerical results
presented in Figs.~\ref{figa2} and~\ref{figa3} satisfy the normalization condition
\begin{eqnarray}
 && \int^\infty_0\d(\Delta E/\omega_c)P(\Delta
E/E,R,\omega_c) \nonumber \\
&& \qquad =p_0(R,m/E) +
  \int^\infty_0\d(\Delta E/\omega_c)p(\Delta E/\omega_c,R,m/E)=1\, .
  \label{a.6}
\end{eqnarray}
Our starting point, the gluon energy distribution (\ref{a.1}), is derived in the
eikonal limit and holds to leading order in $1/E$. In this high energy approximation,
it is possible that for a finite parton energy $E$, the quenching weight has support
in the region $\Delta E > E$. This is clearly an artifact of the formalism. We correct
it by assuming that a parton which is predicted to lose more than its total energy
has lost all its energy in the medium. To this end, we 
truncate $P(\Delta E/E)$ at $\Delta E=E$ but we ensure the normalization
(\ref{a.6}) by adding a $\delta$-function~\cite{Dainese:2004te,Eskola:2004cr}
\begin{eqnarray}
   P_{\rm nrw}(\Delta E/E,R,\omega_c)&=&
    P(\Delta E/E,R,\omega_c)\, \Theta(1 - \Delta E/E) \nonumber \\
    &&+ \delta(\Delta E/E-1)\, \int_1^\infty \d\epsilon\,P(\epsilon)
    \, .
   \label{a.8}
\end{eqnarray}
The corresponding Monte Carlo implementation is: sample an energy loss
$\Delta E$ and set the new parton energy to 0 if $\Delta E\geq E$.

The size of the remainder term $\int_1^\infty \d\epsilon\,P(\epsilon)$
has been used~\cite{Dainese:2004te,Eskola:2004cr} to estimate the systematic 
theoretical uncertainties in this formalism. These uncertainties increase with 
increasing parton energy loss and they decrease with increasing 
transverse momentum. For the nuclear modification factors $R_{AA}$ of
both light-flavored and heavy-flavored hadrons, they amount to typical 
uncertainties of $\pm 0.1$ at $p_T = 5~\gev$
and $\pm 0.05$ at $p_T > 10~\gev$.


\end{document}